\documentstyle[aps,prl,epsf,floats]{revtex}

\begin{document}
\twocolumn[\hsize\textwidth\columnwidth\hsize\csname@twocolumnfalse%
\endcsname

\title{Multichannel pseudogap Kondo model:\\Large-$N$ solution and quantum-critical dynamics}
\author{Matthias Vojta}
\address{Theoretische Physik III, Elektronische Korrelationen und
Magnetismus, 
Universit\"at Augsburg, 86135 Augsburg, Germany}
\date{February 5, 2001} 
\maketitle

\begin{abstract}
We discuss a multichannel SU($N$) Kondo model which displays
non-trivial zero-temperature phase transitions due
to a conduction electron density of states vanishing with a power
law at the Fermi level.
In a particular large-$N$ limit, the system is described by coupled
integral equations corresponding to a dynamic saddle point.
We exactly determine the universal low-energy behavior of spectral
densities at the scale-invariant fixed points,
obtain anomalous exponents, and compute scaling functions
describing the crossover near the quantum-critical points.
We argue that our findings are relevant to recent experiments on
impurity-doped $d$-wave superconductors.
\end{abstract}
\pacs{PACS numbers:}
\vspace*{-18 pt}
]

The Kondo effect in metals is a well studied phenomenon in
many-body physics.
The low-energy physics 
is completely determined by a single energy scale,
the Kondo temperature $T_{\rm K}$,
and the impurity spin is fully quenched in the low-temperature limit,
$T \ll T_{\rm K}$~\cite{hewson}.
The standard picture of the Kondo effect has to be revised
if the conduction band density of states (DOS)
vanishes identically at the Fermi
energy~\cite{withoff,cassa,chen,bulla,GBI}.
This is the case in so-called pseudogap systems
with a power-law DOS $\rho(\epsilon) \sim |\epsilon|^r$ ($r>0$)
which arises in one-dimensional interacting systems,
in certain zero-gap semiconductors,
and in systems with long-range order where
the order parameter has nodes at the Fermi surface, {\em e.g.},
$p$- and $d$-wave superconductors ($r=2$ and 1).
Of special interest 
are high-$T_c$ cuprate superconductors, where indeed non-trivial
Kondo-like behavior has been observed associated with the magnetic
moments induced by non-magnetic Zn impurities~\cite{bobroff}.

The pseudogap Kondo problem has attracted a lot of attention
during the last decade.
A number of studies \cite{withoff,cassa} including the initial work by
Withoff and Fradkin employed a slave-boson large-$N$ technique; 
further progress and insight came from
numerical renormalization group (NRG) calculations \cite{chen,bulla,GBI}
and the local moment approach~\cite{LMA}.
The general picture arising from these studies is that there exists
a zero-temperature phase transition at a critical Kondo
coupling, $J_c$, below which the impurity spin is unscreened
even at lowest temperatures.
Also, the behavior depends sensitively on the presence or absence of
particle-hole (p-h) symmetry: in the p-h symmetric case there
is no complete screening even for $J_{\rm K}>J_c$.
A comprehensive discussion of possible fixed points
and their thermodynamic properties has been given by Gonzalez-Buxton and
Ingersent \cite{GBI} based on the NRG approach.
Spectral properties of the pseudogap Kondo model have so far been
investigated for the p-h symmetric case only~\cite{bulla,LMA},
and the knowledge about the quantum-critical dynamics is
limited.
Importantly, recent advances in scanning tunneling
microscopy (STM) have made it possible to directly
observe local spectral properties of impurities
in superconductors~\cite{seamus}.

The purpose of this paper is to provide an exact low-energy solution of the
pseudogap Kondo problem in a certain large-$N$ limit~\cite{PG0,PG}
which is appropriate for obtaining dynamical properties.
We shall determine anomalous exponents and spectral densities at
the fixed points of interest, and describe the crossovers in the vicinity
of quantum-critical points.
The critical fixed points are found to have exponents varying
continuously with the pseudogap exponent $r$.
To our knowledge, our study represents the first critical theory of
the pseudogap Kondo transition in the presence of p-h asymmetry.


We begin by describing the generalized multichannel Kondo problem,
characterized by $K$ channels of conduction
electrons and a spin symmetry group extended from SU(2) to
SU($N$).
The choice of the SU($N$) representation for the impurity
spin $\hat{S}$ qualitatively influences the phase diagram
in the large-$N$ limit.
For $r=0$, a symmetric (bosonic) representation leads to both
underscreened and overscreened \cite{NB,CZ} phases with an
exactly screened phase in between \cite{PG0},
whereas an antisymmetric representation only allows for an
overscreened situation \cite{PG}.
A second crucial difference is that the bosonic large-$N$
limit contains a generic p-h asymmetry which prevents
a study of a p-h symmetric model.
Therefore, we choose an antisymmetric 
representation \cite{PG} for $\hat{S}$,
corresponding to a Young tableau with a single column of $Q$ boxes.
The $N^2-1$ traceless components of $\hat{S}$
can be written in terms of $N$ auxiliary fermions, $f_{\nu}$, as
$S_{\nu\mu} = f^\dagger_{\nu}f_{\mu}- Q\,\delta_{\nu\mu}/N$,
together with the constraint
$\sum_\nu f^\dagger_{\nu}f_{\nu} = Q$.
The conduction electrons transform under the fundamental
representation of SU($N$), and carry a channel (flavor) index
$i=1,\cdots,K$ and SU($N$) spin index $\nu=1,\cdots,N$.
The Hamiltonian of the model reads:
\begin{equation}
H\, =\, \sum_{{\bf k}i\nu}
\epsilon_{\bf k} c^\dagger_{{\bf k}i\nu} c_{{\bf k}i\nu}
\,+\,J_{\rm K} \sum_{{\bf k}{\bf k'}i\nu\mu}
S_{\nu\mu}\, \, { c^\dagger_{{\bf k}i\mu} c_{{\bf k'}i\nu}}
\label{ham}
\end{equation}
The fermionic bath is assumed to have a power
law DOS at low energies; for simplicity we use a model DOS given by
$\rho(\epsilon) = \rho_0 |\epsilon/D|^r$ for $|\epsilon| < D$.

\begin{figure}
\epsfxsize=3.4in
\centerline{\epsffile{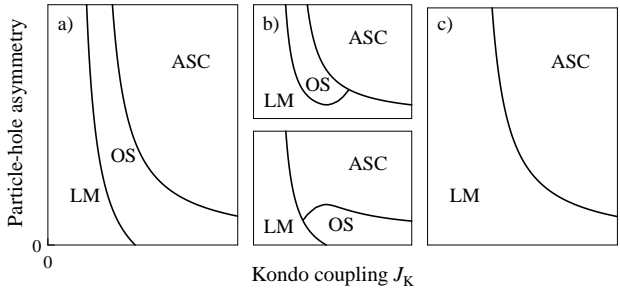}}
\caption{
Schematic phase diagrams of the multichannel pseudogap
Kondo model as function of the Kondo coupling $J_{\rm K}$ and
the particle-hole asymmetry ({\em e.g.} potential scattering),
for pseudogap exponents $r$ with
a) $0 < r < r_{\rm min}^\ast$,
b) $r_{\rm min}^\ast < r < r_{\rm max}^\ast$ for
$\gamma>\gamma_c$ (top) and $\gamma<\gamma_c$ (bottom), $\gamma_c\simeq 0.8$,
c) $r > r_{\rm max}^\ast$.
\vspace{-10pt}
}
\label{figpd}
\end{figure}

For $r=0$, it is known that weak antiferromagnetic coupling $J_{\rm K}$ grows
under renormalization for $K\geq 2$ and any $N$, $Q$, and there
is a stable intermediate-coupling non-Fermi-liquid fixed point
associated with overscreening of the impurity spin.
For $r>0$, we find that the situation is different, and a
number of phases appear:
(i) A local moment (LM) regime is reached for weak initial coupling.
Here $J_{\rm K}$ flows to zero, and the impurity is unscreened.
(ii) An overscreened phase (OS) is found for $0<r<r_{\rm max}^\ast$ and
$J_{c1}<J_{\rm K}<J_{c2}$, it is a natural generalization of the
$r=0$ overscreened fixed point.
(iii) There is a new asymmetric strong-coupling (ASC) phase which
exists for $r>0$, and is reached only for $J_{\rm K} > J_{c2}$ and a
sufficient amount of p-h symmetry breaking.
Qualitative phase diagrams are shown Fig.~\ref{figpd};
the important case of $r=1$ corresponds to Fig.~\ref{figpd}c.

We now proceed with the analysis of the model in the large-$N$ limit
which has been introduced in Refs.~\onlinecite{PG0,PG}.
We set $K=\gamma N$, $J_{\rm K} = J/N$, and take the limit $N\to\infty$
keeping $\gamma$ and $J$ fixed.
The important difference of the present large-$N$ limit to the usual
non-crossing approximation (NCA) \cite{CR} is the treatment
of the occupation constraint: here we set $Q = q_0 N$ and keep
$q_0$ fixed.
The advantage is twofold: we can achieve p-h symmetry with $Q=N/2$,
and in the large-$N$ limit the dynamics of the system is
determined by a true (time-dependent) saddle point~\cite{PG0,PG}.
(Formally, results corresponding to $Q=1$ as in Ref.~\onlinecite{CR}
are obtained here in the limit $q_0\to 0$.)
The Kondo interaction in Eq.~(\ref{ham}) can be decoupled in each channel
by auxiliary bosonic fields $B_i(\tau)$, conjugate to
the amplitude $\sum_\nu c_{i\nu}^\dagger(\tau) f_\nu(\tau)$.
The constraint is implemented with a Lagrange multiplier field
$\lambda(\tau)$.
One finds the following integral equations for the fermionic and
bosonic Green's functions
$G_f(\tau) = -\langle {\rm T}_\tau f_\nu(\tau) f_\nu^\dagger(0) \rangle$,
$G_B(\tau) =  \langle {\rm T}_\tau B_i(\tau) B_i^\dagger(0) \rangle $:
\begin{equation}
\label{sp}
\Sigma_f(\tau)=\gamma G_0(\tau) G_B(\tau)\,\,\,,\,\,\,
\Sigma_B(\tau)= -G_0(-\tau) G_f(\tau) \,,
\end{equation}
with the self-energies $\Sigma_f$ and $\Sigma_B$ defined by:
\begin{eqnarray}
\label{defsigmaf}
G_f^{-1}(i\omega_n) &=& i\omega_n+\lambda-\Sigma_f(i\omega_n)  \\
\label{defsigmab}
G_B^{-1}(i\nu_n) &=& {{1}\over{J}}-\Sigma_B(i\nu_n) \,.
\end{eqnarray}
In these expressions $\omega_n=(2n+1)\pi/\beta$ and $\nu_n=2n\pi/\beta$
denote fermionic and bosonic Matsubara frequencies.
The third saddle point equation fixes the impurity ``occupation'' and thus
determines $\lambda$:
\begin{equation}
\label{eqlambda}
G_f(\tau=0^-) = {1 \over\beta  }
\sum_n G_f (i\omega_n) e^{i\omega_n 0^+}\,=\,q_0
\,.
\end{equation}
Eqs.~(\ref{sp},\ref{defsigmaf},\ref{defsigmab}) are identical in structure
to the usual NCA equations, but the physics is changed by the constraint
eq.~(\ref{eqlambda}) which keeps track of the choice of the impurity
spin representation.
With a symmetric conduction band DOS, the model is p-h symmetric
(invariant under $f^\dagger_{\nu}\leftrightarrow f_{\nu}$ and
$c^\dagger_{{\bf k}i\nu}\leftrightarrow c_{{\bf k}i\nu}$)
for $q_0 = 1/2$.
P-h symmetry can be broken at the impurity spin ($q_0 \neq 1/2$)
as well as in the environment (potential scattering or doping), both
leading to similar effects.

An important dynamic observable is the conduction electron T-matrix
$T(\omega)$ (which corresponds to the impurity
Green's function for an impurity Anderson model), given by the
convolution of $G_f$ and $G_B$, so its Fourier transform is
$T(\tau) = - (1/N) G_{f}(\tau)G_{B}(-\tau)$.
Local susceptibilities for the spin and channel degrees of freedom
are obtained by
$\chi_{\rm loc,sp} = - \int_0^\beta d\tau G_f(\tau) G_f(-\tau)$ and
$\chi_{\rm loc,ch} = \int_0^\beta d\tau G_B(\tau) G_B(-\tau)$;
these have to be distinguished from the impurity contributions
$\chi_{\rm imp,sp,ch}$ to the reponse to a {\em global}
(spin or channel) field.

\begin{figure}
\epsfxsize=3.2in
\centerline{\epsffile{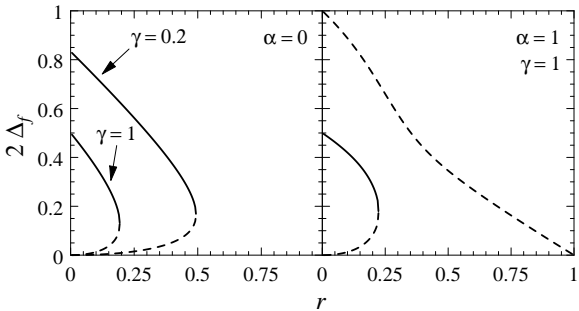}}
\caption{
Values of the exponent $2\Delta_f$ at the
scale-invariant fixed points, obtained from
Eq.~(\protect\ref{expeq}), as function of $r$.
Solid:  infrared stable fixed points (OS).
Dashed: unstable fixed points (quantum phase transitions).
Left: particle-hole symmetric case $\alpha=0$.
Right: with particle-hole asymmetry.
The disappearance of the stable OS solution marks
the value of $r^\ast(\alpha)$.
We find $r_{\rm min}^\ast,r_{\rm max}^\ast \to 1$ for $\gamma\to 0$,
$r_{\rm min}^\ast\simeq 0.19$, $r_{\rm max}^\ast\simeq 0.29$ for $\gamma=1$,
and $r_{\rm min}^\ast,r_{\rm max}^\ast \to 0$ for $\gamma\to\infty$.
(In the single-channel SU(2) case, $r^\ast = \frac{1}{2}$,
whereas $N\!=\!K\!=\!2$ gives $r^\ast\simeq 0.23$ \protect\cite{GBI}.)
\vspace{-10pt}
}
\label{figexp}
\end{figure}

We now turn to a detailed analysis of the low-energy, low-temperature
behavior of the above equations.
In a first step, we analyze possibly existing scale-invariant
fixed points where the Green's functions follow
power laws at small frequencies.
It is important to note that {\em any} continuous quantum phase
transition (as well as other possible intermediate coupling
fixed points) must be described by such behavior.
The constant terms in Eqs.~(\ref{defsigmaf},\ref{defsigmab})
have to vanish,
${\lambda}-\Sigma_f(i0^+)\rightarrow 0, 1/J-\Sigma_B(i0^+)\rightarrow 0$
as $T\to 0$ \cite{PG}.
A power-law ansatz
$G_B(\tau)\sim A_B/\tau^{2\Delta_B}\,,\,
G_f(\tau)\sim A_f/\tau^{2\Delta_f}$ for the Green's
functions in the regime $1/T^*\ll\tau\ll\beta\rightarrow\infty$ (where
$T^*$ is a crossover temperature)
is equivalent to low-energy spectral densities
\vspace{-10pt}
\begin{eqnarray}
\rho_{f,B}(\omega\rightarrow 0^{\pm}) \sim
\frac {A_{f,B,\pm}} {(\pm\omega)^{1-2\Delta_{f,B}}}
\,,
\label{denst0}
\end{eqnarray}
which have the same power-law exponent, but different
prefactors $A_\pm$, for $\omega>0$ and $\omega<0$.
Inserting these Green's functions into Eqs.~(\ref{sp},\ref{defsigmaf},\ref{defsigmab})
one finds the following equations for the exponents:
\begin{eqnarray}
2\Delta_f+2\Delta_B &=& 1-r \,,
\nonumber \\
\gamma
\frac { (1-2\Delta_B) \sin 2\pi\Delta_B}
      {\sin^2 \pi \Delta_B + \sinh^2 \alpha/2}
&=&
\frac {(1-2\Delta_f) \sin 2\pi\Delta_f}
      {\cos^2 \pi \Delta_f + \sinh^2 \alpha/2}
\,.
\label{expeq}
\end{eqnarray}
Here, $\alpha$ is a spectral asymmetry parameter defined
by $\exp(\alpha) = A_{f,-} / A_{f,+} = -A_{B,-} / A_{B,+}$.
For $r=0$, $\alpha$ drops out,
and there is a single solution with non-trivial exponents,
$2\Delta_f = 1/(1+\gamma)$, $2\Delta_B = \gamma/(1+\gamma)$,
which describes an infrared stable fixed point corresponding
to an overscreened spin \cite{PG}.
In contrast, the behavior for $0<r\leq 1$ is much richer, and is
illustrated in Figs.~\ref{figpd} and \ref{figexp}.
For small $r$ and $\alpha\neq 0$, one finds three solutions,
one infrared stable representing the OS phase,
and two unstable corresponding to the quantum phase
transitions from OS to LM and ASC.
With increasing $r$, two solutions disappear at $r^\ast(\alpha)$,
leaving a single unstable solution for the direct
transition between the LM and ASC phases.
In the p-h symmetric case ($\alpha$=0), the
unstable solution for the transition to ASC is absent.
Therefore, no scale-invarian solution is found for $r>r^\ast(0)$:
the p-h symmetric system here always flows to the LM weak
coupling phase.
The critical value $r^\ast(\alpha)$ is bounded,
$r_{\rm min}^\ast < r^\ast < r_{\rm max}^\ast$.
From the presence or absence of intermediate coupling fixed
points one can infer phase diagrams as in Fig.~\ref{figpd};
these are also well borne out by our numerical
studies of Eqs.~(\ref{sp},\ref{defsigmaf},\ref{defsigmab}).
In the narrow range $r_{\rm min}^\ast < r < r_{\rm max}^\ast$,
the existence of the OS phase depends on the amount of
p-h symmetry breaking, Fig.~\ref{figpd}b.

The $N\!=\!K\!=\!2$ version of the overscreened pseudogap Kondo
model has been studied earlier by NRG \cite{GBI}, with results
somewhat different from the large-$N$ limit.
The $N\!=\!K\!=\!2$ model has a single $r^\ast$ value,
no ASC region for $r < r^\ast \simeq 0.23$, and
a minimal potential scattering value to reach ASC
for $r > r^\ast$.
Some of these differences arise from the fact
that the $N\!=\!K\!=\!2$ case exhibits isolated critical fixed points,
whereas the scale-invariant large-$N$ solutions actually correspond to
continuous {\em lines of fixed points}, 
parametrized by the low-energy spectral asymmetry $\alpha$,
which is determined by the symmetry-breaking parameters of
the model.


The universal low-energy behavior of the T-matrix at the scale-invariant
fixed points is readily obtained from the convolution of $G_f$ and $G_B$,
its spectral density $\rho_{\rm T}(\omega\rightarrow 0)$ diverges as
$A_{\rm T} |\omega|^{-r}$.
Notably, the prefactors for both signs of $\omega$ are equal here, {\em i.e.},
the asymmetry described by $\alpha$ disappears in the scaling limit \cite{PG}.
Furthermore, the amplitude $A_{\rm T}$ in the OS phase is independent of
$J$, this is the analog of Friedel's sum rule. 
The scaling forms of the {\em finite-temperature} spectral densities
can be determined \cite{PG}
from the $T=0$ quantities (\ref{denst0}) by applying a conformal
transformation $z = \exp(i2\pi\tau/\beta)$.
Low-temperature thermodynamic quantities can now be calculated,
{\em e.g.}, the local spin susceptibility \cite{PG,CR} follows
$\chi_{\rm loc,sp} \sim 1/T^{1-4\Delta_f}$ for $4\Delta_f<1$,
$\chi_{\rm loc,sp} \sim \ln (1/T)$ for $4\Delta_f=1$, and
$\chi_{\rm loc,sp} \sim {\rm const}$ otherwise; similar
results hold for $\chi_{\rm loc,ch}$ in terms of $\Delta_b$.

In a second step, we discuss possible non-scale-invariant
fixed points of the system (which are absent at $r=0$):
the local moment (LM) and the asymmetric strong coupling
(ASC) fixed point.
These solutions of Eqs.~(\ref{sp},\ref{defsigmaf},\ref{defsigmab})
appear if the $T=0$ ``compensation'' of the constant terms in
Eqs.~(\ref{defsigmaf},\ref{defsigmab}) is not realized.

For the LM fixed point, reached
for small Kondo coupling, it is important to notice
that $\Sigma_B(i0+)$ is bounded from above for $r>0$
and cannot compensate large $1/J$ values
(the corresponding integral is infrared divergent for $r=0$).
Hence, there is no power-law solution for $J$ smaller than a critical $J_{c1}$.
A constant term in $G_B^{-1}$ implies a $\delta$-peak in the $T=0$
fermion spectral density; furthermore the spectral asymmetry
flows to zero in the scaling limit, leading to
$\rho_f(\omega) \sim \delta(\omega)$.
This is equivalent to an unscreened spin:
the local spin susceptibility follows a Curie law for $T\to 0$.
The ``local moment'' $T\chi_{\rm loc,sp}$ is reduced from
its free spin value, but $T\chi_{\rm imp,sp}$, measuring
the {\em total} impurity-induced Curie contribution to the uniform
susceptibility, is that of a free spin, {\em i.e.},
part of the moment is carried by the surrounding
conduction electrons due to the residual coupling
mediated by $G_B$.

The ASC fixed point is reached at sufficient particle-hole asymmetry and
Kondo coupling.
Here, ${\lambda}-\Sigma_f(i0^+)$ does not vanish for $T\to 0$.
This leads to a $\delta$-peak in the auxiliary boson
spectral density, $\rho_B(\omega) \sim \pm \delta(\omega-0^\pm)$
[the sign depending on ${\rm sgn}(q_0-1/2)$~],
which implies maximal spectral asymmetry in the scaling limit.
The channel susceptibility diverges like $1/T$, and the system has
a free ``moment'' in the channel (flavor) degrees of
freedom, also leading to a residual \mbox{$T=0$} entropy.
In contrast, the spin is completely screened,
$\chi_{\rm imp,sp}$, $\chi_{\rm loc,sp} \to {\rm const}$.
The ASC phase can be understood as $(N\!-\!Q)$ conduction electrons
bound to the impurity;
for $N\!=\!K\!=\!2$ this is equivalent to a spin singlet,
channel doublet state~\cite{GBI}.

For both the LM and ASC fixed points, the spectral density
$\rho_{\rm T}(\omega)$ of the T-matrix behaves as $|\omega|^{r}$ at small
frequencies, and the spectral asymmetry disappears in the scaling
limit of $\rho_{\rm T}$.

At this point we briefly comment on the case $r\geq 1$:
the OS phase disappears for all $\gamma$;
in the presence of particle-hole asymmetry a transition between
LM and ASC is still possible.
For $r=1$ the above power law analysis remains valid, but has to be
supplemented with logarithmic corrections.
For \mbox{$r>1$} the critical power law at the LM-ASC transition is
replaced by $\rho_{\rm T} \sim |\omega|^{r-2}$.

After having described the fixed points
we now turn to the universal behavior near the quantum phase transitions.
In the vicinity of each critical point, one can define an energy scale
$T^*$, which vanishes at the transition, and defines the crossover
energy above which quantum-critical behavior is observed~\cite{book}.
From the above low-energy analysis one can deduce the dependence
of $T^*$ on the reduced coupling, $j= (J-J_c)/J_c$, measuring the distance
from the critical point at $J_c$.
Approaching the LM-OS or LM-ASC transition from small $J$ we have
$T^* \sim (-j)^{1/r}$, from large $J$ we find $T^* \sim j^{1/(r+2\Delta_{f,\rm cr})}$
where $\Delta_{f,\rm cr}$ is the anomalous fermion exponent at the critical
fixed point.
Similarly, we have $T^* \sim (-j)^{1/(r+2\Delta_{f,\rm OS})}$ and
$T^* \sim j^{1/(r+2\Delta_{f,\rm cr})}$ if we approach the OS-ASC transition
from below or above, respectively.
The local susceptibilities are easily derived, {\em e.g.},
in the ASC phase we have
$\chi_{\rm loc,sp}(T$=0) $ \sim T^{*(4\Delta_{f,\rm cr}-1)} \sim j^{-\bar{\gamma}}$
with $\bar{\gamma} = (1-4\Delta_{f,\rm cr})/(r+2\Delta_{f,\rm cr})$.

\begin{figure}
\epsfxsize=2.9in
\centerline{\epsffile{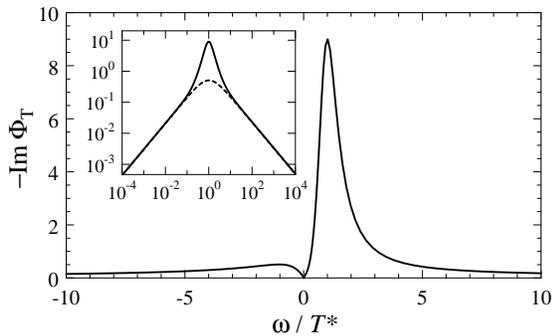}}
\caption{
Zero-temperature scaling function $\Phi_{\rm T}$
for the conduction electron T-matrix at $r=0.8$,
$\gamma=1$, $q_0=0.1$,
describing the crossover from the ASC behavior
at low energies to quantum criticality
at high energies.
Inset: Same on log-log scale, showing both
$\bar{\omega}>0$ (solid) and $\bar{\omega}<0$ (dashed).
\vspace{-10pt}
}
\label{figscl1}
\end{figure}

Near a critical point all observables show
scaling behavior as function of $T/T^*$ and $\omega/T^*$ ($k_B=1$).
For instance, the 
T-matrix obeys the scaling form
\vspace{-8pt}
\begin{equation}
T(\omega) = \frac{{\cal A}}{T^{*(1-\eta_{\rm T})}} \,
\Phi_{\rm T} \! \left(
\frac{\omega}{T^*}, \frac{T}{T^*} \right)
\,,
\label{gscale}
\end{equation}
where ${\cal A}$ is an amplitude prefactor, $\eta_{\rm T}=1-r$ is
the anomalous exponent, and $\Phi_{\rm T}$ is a universal scaling
function (for the particular transition and values of $r$,
$\gamma$, $q_0$).
As an example, we focus on the transition between the LM and ASC
phases, for $J>J_c$.
A result for the $T=0$ scaling function $\Phi_{\rm T}$ is shown in
Fig.~\ref{figscl1}.
For $\bar{\omega} = \omega/T^* \ll 1$ we have ASC behavior with
${\rm Im} \Phi_{\rm T} \sim |\bar{\omega}|^r$,
for $\bar{\omega} \gg 1$ the spectral density follows the
quantum-critical power law $|\bar{\omega}|^{-r}$
as predicted above.
In both limits the spectrum is particle-hole symmetric (!), however,
in the crossover region the asymmetry leads to a large
peak for one sign of $\bar{\omega}$
[depending again on ${\rm sgn}(q_0-1/2)$~].

Finally, we mention possible applications of our results.
Although we have taken a multichannel large-$N$ limit here,
the T-matrix spectral properties of the LM, ASC, and critical fixed points
turn out to apply also to the single-channel case~\cite{bulla2}.
In most systems there is a finite particle-hole asymmetry, and if Kondo
screening is present in a pseudogap system (and $r$ is not too small),
the physics will be described by the ASC fixed point~\cite{GBI}.
Then, in contrast to the metallic $r=0$ case, the impurity spectral
function does not show a peak {\em at} the Fermi level, but at an energy
of order $T_{\rm K}\sim T^*$ (Fig.~\ref{figscl1}, see also \cite{cassa}).
Interestingly, STM studies of Zn impurities in the cuprate $d$-wave
superconductor BSCCO \cite{seamus} have shown a large peak
in the differential conductance
at a small bias of 1.5 meV, and a likely possibility is
that this peak arises from the Kondo screening of the impurity-induced
moment \cite{tolya}.
The value of $T^*$ = 15 K is consistent with NMR studies \cite{bobroff}
which show a $T_{\rm K}$ of order 20--40 K around optimal doping.
Last not least, genuine multichannel pseudogap physics may be found
in quantum dots coupled to $d$-wave-superconducting leads,
and possibly in heavy-fermion superconductors \cite{CZ}.


The author thanks H. Alloul, R. Bulla, A. Georges,
O. Parcollet, A. Polkovnikov, T. Pruschke, and S. Sachdev
for invaluable discussions.
This research was supported by the DFG through SFB 484.

\end{document}